\documentclass[12pt]{article}

\usepackage{amsfonts,amscd,amssymb}
\usepackage[matrix,arrow,frame,curve]{xy}
\CompileMatrices
\def\be{\begin{displaymath}}
\def\ee{\end{displaymath}}

\begin{document}
\textwidth 160mm
\textheight 240mm
\topmargin -20mm
\oddsidemargin 0pt
\evensidemargin 0pt

\newcommand{\beq}{\begin{equation}}
\newcommand{\eeq}{\end{equation}}
\newcommand{\CC}{\mathbb C}
\def\AA{\mathop{\mathbb A}\nolimits}
\def\TT{\mathop{\mathbb T}\nolimits}
\def\ZZ{\mathop{\mathbb Z}\nolimits}
\def\PP{\mathop{\mathbb P}\nolimits}

\def\a{\alpha}
\def\pa{\partial}
\def\sn{\mathop{\rm sn}\nolimits}
\def\cn{\mathop{\rm cn}\nolimits}
\def\dn{\mathop{\rm dn}\nolimits}
\def\cd{\mathop{\rm cd}\nolimits}
\def\dc{\mathop{\rm dc}\nolimits}


\begin{flushright}
ITEP-TH-41/01\\
LPTHE-01-62 \\
hep-th/0111066
\end{flushright}

\setcounter{footnote}{0}
\centerline{\bf Double-Elliptic Dynamical Systems }
\bigskip
\centerline{\bf from Generalized
Mukai-Sklyanin  Algebras}
\vspace{0.5cm}

\centerline{H. W. Braden$^{1}$, A. Gorsky$^{2,3}$, A. Odesskii$^{4}$,
V. Rubtsov$^{2,5}$}

\bigskip
\begin{center}
$^{1}$ {\em Department of Mathematics and Statistics,
The University of Edinburgh, Edinburgh, UK}\\
$^{2}$ {\em Institute of Theoretical and Experimental Physics, B.
Cheryomushkinskaya 25, Moscow}\\
$^{3}$ {\em LPTHE, Universite Paris VI,
4 Place Jussieu, Paris, France}\\
$^{4}$ {\em Landau Institute for Theoretical Physics, Kosygina 2, Moscow}\\
$^{5}$ {\em Department de Mathematiques, Universite d'Angers,
2, bd. Lavoisier, 49045, Angers}
\end{center}

\vspace{2.0cm}

\begin{abstract}
We consider the double-elliptic generalisation of
dynamical systems of Calogero-Toda-Ruijsenaars
type using  finite-dimensional Mukai-Sklyanin algebras.
The two-body system, which involves an
elliptic dependence both on coordinates and momenta,
is investigated in detail and the relation with Nambu
dynamics is mentioned.
We identify the 2D complex manifold associated with the double elliptic system
as an elliptically fibered rational ("1/2K3 ") surface.
Some  generalisations
are suggested  which provide the ground for a description
of the N-body systems.  Possible applications to
SUSY gauge theories with adjoint matter in $d=6$
with two compact dimensions are
discussed.
\end{abstract}
\vspace{2.0cm}
\section{Introduction}

Hamiltonian systems of both Calogero-Toda-Ruijsenaars
type and spin chains
can naturally be associated with finite or infinite
dimensional groups via Hamiltonian reduction.
The Hamiltonian reduction procedure produces a
unified description of such systems in terms of
free motion on the group-like symplectic manifold
restricted to a fixed level of the moment map \cite{kks}. The line
of generalisation from rational to trigonometric to elliptic models
has a clear algebraic counterpart in terms of zero loop, one-loop and two-loop
structures \cite{gn}. One further has to distinguish two
different possible generalisations, where either coordinates or momenta
take values on the elliptic curve. It is useful to introduce
the notion  of ``dual" system here \cite{fgnr, bmmm}, where the
coordinates and action variables effectively get interchanged.
The attempt to give a mathematically rigorous definition to the duality
of integrable systems has been recently given in \cite{HausTadd}.

The line of generalisation in terms of increasing the number of loops to
produce say the trigonometric Ruijsenaars models from the rational ones
is not the only one that has been considered.
The same results can be obtained for the rational and trigonometric
Ruijsenaars models moving in another direction \cite{fgnr}. Namely, there it
was shown that these models admit the realisation as motion on a finite
dimensional quantum group. However a similar interpretation for the elliptic
models was lacking. Below we shall suggest a
realisation of the elliptic models as motion associated with certain finite
dimensional algebras.\footnote{The preliminary results of this
paper were presented in the review \cite{gr}.}
It seems that the proper algebraic object for the
general double-elliptic system is the generalised Mukai-Sklyanin
algebra. This was introduced for two quadrics in $\CC^4$ in \cite{sklyanin}
and may be easily generalised to $n$ quadrics in $\CC^{n+2}$ in such a way
that the famous Mukai symplectic Poisson brackets are obtained by considering
the corresponding $K3$ surfaces as affine symplectic leaves of the $n$
quadric structure \cite{mukai, OR}.

In this paper we shall use the algebra induced by four
quadrics in $\CC^6$ which yields the phase space for the two-body
double-elliptic problem.  This model was given explicitly in \cite{bmmm}
in Darboux variables. We shall show that this formulation
of the double elliptic system is equivalent to the
one suggested in \cite{fgnr}.
All other two-body elliptic models such as the elliptic
Ruijsenaars and Calogero models, as well as the periodic Toda chain,
can be derived from the double-elliptic
system via one or other degeneration. Recall that the elliptic
Calogero system is an example of a Hitchin system \cite{Hitchin}.
More general integrable  systems corresponding to Hitchin systems for
curves with many marked points were discussed
in \cite{ER, Nekr, HurtMark}.

Of course the real problem to be solved is a derivation
of the integrable many-body system. In the relatively simple
cases known so far, one could just consider a higher rank group,
which can be finite or infinite dimensional. This program works
perfectly up to the elliptic Ruijsenaars model which is
the most complicated system treated
successfully in this way \cite{aru}. However
the next step to the double-elliptic system appears to be
very complicated technically  and so one has to look for another
technique. One  method, which suggests a complementary description,
is the separation of variables. This approach yields a
parametrization of the phase space of the many-body system as
the symmetric power of some complicated two dimensional
complex manifold \cite{gnr}.
The most difficult part  of analysis in this picture
is the proper identification of the two-dimensional complex
manifold corresponding to the two-particle problem.

The problem of a geometrical description of the
double-elliptic systems, as well as being  intrinsically interesting
from the point of integrability, has important applications to the
description of low-energy effective actions in SUSY
gauge theories in different dimensions.
Via Seiberg-Witten theory it is known that the classical dynamics
of integrable many-body systems addresses quantum issues in SUSY gauge
theories (see \cite{gm, BKr} and the references therein for a review).
The most complicated SUSY systems treated in this manner thus far have
been the elliptic Ruijsenaars model, corresponding
to a D=5 SUSY theory with adjoint matter
and with one compact dimension\cite{nikita5, bmmm2},
and theories with fundamental matter in D=5, 6 \cite{ggm}.
The double-elliptic system supplies the next step to a D=6 SUSY system
with adjoint matter, and where two dimensions
are compact. The two elliptic moduli to be
found in the gauge theory can be identified
with the complexified coupling constant
and the ratio of the radii in the fifth and sixth
dimensions. Therefore the answers we obtain will be of the immediate
use for this theory. The pure D=6 SUSY theory
with two compact dimensions corresponds
to the degeneration of the double-elliptic
system to the ``elliptic Toda'' one.

Let us also note that a clear representation
of the integrable systems corresponding to
D=6 SUSY gauge theories is also necessary
to formulate the dualities from \cite{fgnr,bmmm}
precisely. The point is that these dualities relate gauge theories in
different dimensions. For instance, the systems dual
to the elliptic Calogero or periodic Toda system (D=4)
correspond to D=6 gauge
models. Another example involves the
relationship between Ruijsenaars (D=5) models
and the trigonometric Calogero (D=4) ones.
It seems that this duality will be very helpful in the nonperturbative
(De)-construction of dimensions recently discussed in \cite{csaki}.

The paper is organised as follows. We start with general remarks
concerning the generalised Mukai-Sklyanin algebra. Then we identify the
four-dimensional manifolds responsible for the elliptic models
including the double elliptic one. The possible
generalisation to the $n$-body
problems will be suggested. Finally we apply
the double-elliptic system  to the
description of the
$SU(2)$ SUSY gauge models with the
adjoint matter in D=6 where two dimensions
are assumed to be compact.

\section{Two-body system}

\subsection{General facts}
The generalised Mukai-Sklyanin algebra provides a  Poisson  structure on the
intersection of $n$ polynomials  $Q_i$ in ${\CC}^{n+2}$. The case of $n=2$ quadrics
corresponds to the Sklyanin algebra while the case $n=3$ corresponds to the Poisson
structure attributed to a K3 surface. The Poisson bracket in affine coordinates looks
as follows: \beq \{x_i,x_j\}= \epsilon_{ij k_1 \dots k_n}\frac{\partial Q_1}{\partial
x_{k_1}} \dots \frac{\partial Q_n}{\partial x_{k_n}}, \eeq where $x_i$ are affine
coordinates in ${\CC}^{n+2}$. The corresponding four dimensional manifold is ``Poisson
noncommutative", a point we shall amplify in due course. The polynomials $Q_i$
themselves yield the Casimirs of the algebras.

We choose the following system of four quadrics in ${\CC}^6$ which
provides the phase space for the two-body double-elliptic system
\beq
\label{quads1}
\begin{array}{rcl}
\label{quad}
x_1^2 - x_2^2 =1 \\
x_1^2 - x_3^2 =k^2 \\
-g^2x_1^2 + x_4^2 + x_5^2=1 \\
-g^2  x_1^2 +  x_4^2 +  \tilde {k}^{-2}x_6^2=\tilde {k}^{-2}.
\end{array}
\eeq
Here the first pair of equations yields the ``affinization" of the
projective embedding of the elliptic curve into ${\CC}P^{3}$, while
the second pair provides the elliptic curve  which locally is
fibered over the first elliptic curve. If the coupling constant
$g$ vanishes the system is just  two copies of elliptic curves
embedded in ${\CC}^{3}\times {\CC}^{3}$. Let us emphasise that
the coupling constant amounts to an additional noncommutativity
between the coordinates beyond the standard noncommutativity
of coordinates and momenta.

We should also warn that we will use the same letters to denote the
homogeneous and non-homogeneous coordinates in all non-confusing
cases.

\subsection{Poisson structure}
The relevant Poisson brackets for this particular system of quadrics reads
$$
\begin{array}{c}
\{x_1,x_2\}= \{x_1,x_3\}=\{x_2,x_3\}=0, \\
\{x_5,x_1\}= - x_2 x_3 x_4 x_6, \qquad
\{x_5,x_2\}= - x_1 x_3 x_4 x_6, \qquad
\{x_5,x_3\}= - x_1 x_2 x_4 x_6, \\
\{x_5,x_4\}= - g^2 {\tilde k}^{-2}2 x_1 x_2 x_3 x_6, \qquad
\{x_5,x_6\}=0.
\end{array}
$$
(Various factors of two have been incorporated into the definitions of the
quadrics to avoid unnecessary factors in these Poisson brackets.)
We should remark that the Poisson structure is singular and can't be
extended up to a holomorphic structure on the whole ${\CC}P^{6}$ because of
the arguments of \cite{OR}. Namely, let $X_1,...,X_n$ be coordinates on
${\CC}^n$ considered as an affine part of the
corresponding projective space ${\CC}P^n$ with the homogeneous coordinates
$\left(x_0:x_1:\cdots :x_n\right), X_i = \frac{x_i}{x_0}$. Then if
$\{X_i,X_j\}$ extends to a holomorphic Poisson structure on ${\CC}P^{n}$ the
maximal degree of the structure (= the length of monomials in $X_i$) is $3$ and
\begin{equation}
\label{holomorh}
X_k\{X_i,X_j\}_3 + X_i\{X_j,X_k\}_3 + X_j\{X_k,X_i\}_3 = 0, i\neq j
\neq k.
\end{equation}
Thus  $\{X_i,X_j\}_3 = X_iY_j -  X_jY_i$, with  $deg Y_i = 2$.

The nontrivial commutation relations between coordinates on the distinct tori
correspond to the standard phase space Poisson brackets while the
nontrivial bracket $\{x_5,x_4\}$ is the additional noncommutativity
of the momentum space mentioned earlier.

\subsection{Equations of motion}

Let us note that the triple $x_1, x_2, x_3$ can be considered in the elliptic
parametrization
\begin{eqnarray}
\label{dell1}
&x_1=\frac{1}{sn(q|k)}\\
&x_2=\frac{cn(q|k)}{sn(q|k)}\\
&x_3=\frac{dn(q|k)}{sn(q|k)}
\end{eqnarray}
with the other three coordinates of (\ref{quad}) expressed by the
natural Jacobi functions uniformizing the second (``momenta'') elliptic
curve:
\begin{eqnarray}
\label{dell2}
&x_4=\alpha sn(\beta p|{\frac{\tilde k \alpha}{\beta}})\\
&x_5=\alpha cn(\beta p|{\frac{\tilde k \alpha}{\beta}})\\
&x_6=\beta  dn(\beta p|{\frac{\tilde k \alpha}{\beta}}).
\end{eqnarray}
Here
\beq
\label{alpha} \alpha(q|k)=\sqrt{1+\frac{g^2}{sn^2(q|k)}},
\eeq
\beq
\label{beta} \beta(q|k,\tilde{k})=\sqrt{1+\frac{g^2\tilde{k}}{sn^2(q|k)}},
\eeq

The Hamiltonian of the double-elliptic system in the form of \cite{bmmm}
\beq
\label{ham}
H(p,q)=\alpha(q|k)\,cn(p \beta(q|k,\tilde{k})| \frac{\tilde{k} \alpha(q|k)}{
\beta(q|k) })
\eeq
coincides\footnote{We have chosen a value of the coupling constant
$g$ to be pure imaginary and rescaled by $\sqrt{2}$ with respect to the
choice of \cite{bmmm}.} with $x_5$.

We can check directly (see the Appendix) that the canonical  equations of
motion can be re-written as the following polynomial system
\begin{equation}
\begin{array}{rcl}
{\dot x_1}=\{x_1,x_5\}= x_2x_3x_4x_6\\
{\dot x_2}=\{x_2,x_5\}= x_1x_3x_4x_6\\
{\dot x_3}=\{x_3,x_5\}= x_1x_2x_4x_6\\
{\dot x_4}=\{x_4,x_5\}= g^2x_1x_2x_3x_6\\
{\dot x_5}=\{x_5,x_5\}\equiv 0\\
{\dot x_6}=\{x_6,x_5\}=0.
\label{dellpol}
\end{array}
\end{equation}

Due to the commutation relations $x_6$ is a constant, $x_6=c$ say,
which has to be related to the energy $x_5 = E$ to provide the consistency
of the system which comes from the last two equations of (\ref{quads1}):
\beq \tilde{k}^2 (1-E^2)=1-c^2. \eeq
This consistency condition reduces the number of quadrics in the system
because of the coincidence of the last two equations.

To get an explicit
form of the spectral curve which corresponds to the solution to the equations
of motion let us put $y =x_1x_2x_3x_4$ and obtain the following equation
\beq
\label{spectral}
y^2 = x_1^2(x_1^2 - 1)(x_1^2 - k^2 )(g^2 x_1^2 -( E^2 -1)).
\eeq
Taking $s = x_1^2$ and assuming the energy level to be ``generic"
($E^2 \neq 1,g^2 + 1, g^2 k^2 + 1$)
we obtain that the curve under consideration is a
two-sheeted covering of the following hyperelliptic form of an elliptic curve
$$
y^2 = s(s - 1)(s - k^2 )(g^2 s - (E^2 - 1))
$$
with the branching points $0, 1, k^2,(E^2 -1)/g^2$. We can immediately
integrate the system by the hyperelliptic integral:
$$
dt = {2 x_1{dx_1}\over{x_1x_2 x_3 x_4}} = {2x_1{dx_1}\over{y}} ={{ds}\over{y}} ,
$$
$$
t = \int {{ds}\over{y}} =2 \int{{dx_1}\over{\sqrt{(x_1^2 - 1)(x_1^2 -
k^2 )(g^2 x_1^2 -(E^2 - 1))}}}.
$$

{\bf Remark 1.} Let us suppose that the coupling constant $g$ in the
hamiltonian (\ref{ham}) is equal to $0$ (the case of ``free'' motion).
Then the expression of (\ref{ham}) reduces to
\beq
\label{freeham}
H(p,q) = cn(p|\tilde k)
\eeq
and the parametrization of the second curve is simplified as
\begin{eqnarray}
\label{free}
&x_4= sn(p|{\tilde k})\\
&x_5= cn(p|{\tilde k})\\
&x_6= dn(p|{\tilde k}).
\end{eqnarray}

We have the following ``decoupled'' family of the quadrics in
${\CC}^{3}\times {\CC}^{3}$:
$$
\begin{array}{rcl}
\label{quadfree}
 x_1^2 - x_2^2 =1 \\
 x_1^2 - x_3^2 =k^2 \\
 x_4^2 + x_5^2 =1 \\
\tilde {k}^2 x_4^2 + x_6^2 =1.
\end{array}
$$

The geometrical picture of this family is of course an affine part of the
direct product of the elliptic curves ${\cal E}_{q}(k)$ parametrized by
$x_1,x_2,x_3$ and  ${\cal E}_{p}(\tilde k)$ parametrized by
$x_4,x_5,x_6$.

The hamiltonian again is identified with $x_5$ and we obtain a new
involutive coordinate $x_4$. The canonical equations of the ``free''
motion may be written down immediately:
$$
\begin{array}{rcl}
{\dot x_1}=\{x_1,x_5\}= x_2x_3x_4x_6\\
{\dot x_2}=\{x_2,x_5\}= x_1x_3x_4x_6\\
{\dot x_3}=\{x_3,x_5\}= x_1x_2x_4x_6\\
{\dot x_4}=\{x_4,x_5\}= 0\\
{\dot x_5}=\{x_5,x_5\}\equiv 0\\
{\dot x_6}=\{x_6,x_5\}=0.
\end{array}
$$
and after restriction on a level $x_5 = E$ we can easily recognise in
the above system the classical Arnold-Euler-Nahm top for $SU(2)$ in
the variables $x_1,x_2, x_3$.

{\bf Remark 2.}
Let us consider the ``singular energy level'' $E = \frac{+}{}1$. Then
we have the following nodal rational curve
$$
y^2 = g^2 s^2(s - 1)(s - k^2 )
$$
which may be reduced to the following canonical form
$$
Y^2 = g^2 (1- S)(1- k^2 S)
$$
by the transformation
$$
S = 1/s,\ Y = y/s^2.
$$

{\bf Remark 3.}
We note that the spectral curve of the double
elliptic system (\ref{spectral}) has a clear
physical meaning from the point of view
of D=6 gauge theory. It is this curve
that provides the information concerning the
low energy effective action as well as
the spectrum of BPS states. The energy
level E corresponds to the coordinate on the Coulumb branch
of the moduli space in the corresponding gauge theory. The masses of the
BPS particles can be derived calculating the integrals of the meromorphic
differentials over the cycles on the spectral curve.
The singular values of the energy, at which points the curve degenerates,
correspond to the singularities on the Coulumb branch of the moduli space when
the mass of some BPS particle vanishes.

\subsection{Another choice of the double-elliptic ansatz}

 The choice of another solution for the ansatz of \cite{bmmm} should give
an another reparametrisation for the double-elliptic phase space. It is clear
that all such choices are related to the initial by an appropriate modular
transformation of the Jacobi moduli $k,\tilde k, k_{eff}, \tilde k_{eff}$
(see \cite{bmmm}).  In fact, we are able to check the following statement:
The second choice of the double-elliptic ansatz in \cite{bmmm}
provides us with a polynomial Hamilton system which is equivalent to
the first.

Indeed, the Hamiltonian of the double-elliptic system in the form of the second
solution of \cite{bmmm} is
\beq H(p,q)=\alpha(q|k)\, cn(p|\tilde{k} \alpha(q|k)) \eeq
where now $\alpha(q|k)=\sqrt{1 + \frac{g^2}{cn^2(q|k)}}$.
Taking the the following parametrizations of the coordinate and momentum tori
$$
\begin{array}{c}
x_1=\frac{1}{cn(q|k)},\qquad
x_2=\frac{sn(q|k)}{cn(q|k)},\qquad
x_3=\frac{dn(q|k)}{cn(q|k)};
\\
y_1=sn(p|\tilde k\alpha(q,k)),\qquad
y_2=cn(p|\tilde k\alpha(q,k)),\qquad
y_3=dn(p|\tilde k\alpha(q,k))
\end{array}
$$
and then denoting by
$$
\begin{array}{c}
x_4 = y_3,\qquad
x_5 = \sqrt{1 + g^2x_1^2}\, y_1,\qquad
x_6 = \sqrt{1 + g^2x_1^2}\, y_2
\end{array}
$$
we obtain the four quadrics in ${\CC}^6$:
$$
\begin{array}{rcl}
x_1^2 - x_2^2 =1 \\
x_1^2 - x_3^2 =k^2x_2^2 \\
-g^2x_1^2 + x_5^2 + x_6^2 = 1\\
{\tilde k}^2 x_5^2 + x_4^2 = 1.
\end{array}
$$

Again the double-elliptic  hamiltonian coincides with $x_5$ and the system
associated with the quadrics has a form similar to (\ref{dellpol}):
$$
\begin{array}{rcl}
{\dot x_1}=\{x_1,x_5\}= -x_2x_3x_4x_6\\
{\dot x_2}=\{x_2,x_5\}= -x_1x_3x_4x_6\\
{\dot x_3}=\{x_3,x_5\}= -{k'}^2x_1x_2x_4x_6\\
{\dot x_6}=\{x_6,x_5\}= -g^2x_1x_2x_3x_4\\
{\dot x_5}=\{x_5,x_5\}\equiv 0\\
{\dot x_4}=\{x_4,x_5\}=0.
\end{array}
$$
Here ${k'}^2 +k^2 =1$ is the complementary modulus and the variable $x_4$
plays the same role as $x_6$ in the case (\ref{dellpol}) and also can be
eliminated to obtain an almost identical polynomial form.
As was argued in \cite{bmmm}, this choice of solution better the study
of degenerations of the model (see below chapter 4).

\section{Duality}

Let us now describe the ``duality" property
for the dynamical system under consideration. We shall formulate
the duality as an involution of the four dimensional
manifold. First  perform the following involution of homogeneous coordinates
\beq
x_1\rightarrow x_0,\qquad x_0\rightarrow x_1,\qquad x_2\rightarrow x_5
\eeq
which yields the following intersection of four quadrics
$$
\begin{array}{rcl}
 x_0^2 - x_5^2 = x_1^2\\
 x_0^2 - x_3^2 =k^2 x_1^2 \\
 -g^2x_0^2 + x_4^2 + x_2^2-x_1^2=0 \\
 -g^2 \tilde {k}^2 x_0^2 + \tilde {k}^2 x_4^2 +  x_6^2=x_1^2.
\end{array}
$$

The dual Hamiltonian of \cite{bmmm} is
\beq
\label{hamdual}
H_{dual}= cn(q|k) = \frac{x_2}{x_1}
\eeq
In terms if the affine coordinates $z_i, 1\leq\ i\leq\ 6$:
$$
z_1 = {\frac{x_0}{x_1}},z_2 = {\frac{x_5}{x_1}},z_3 = {\frac{x_3}{x_1}},z_4 =
{\frac{x_4}{x_1}},z_5 = {\frac{x_2}{x_1}},z_6 = {\frac{x_6}{x_1}}.
$$
our ``dual" system of quadrics has ``affine" form
$$
\begin{array}{rcl}
 z_1^2 - z_5^2 = 1\\
 z_1^2 - z_3^2 ={\tilde k}^2\\
 -g^2z_1^2 + z_4^2 + z_2^2 = 1\\
 -g^2 k^2 z_1^2 + k^2 z_4^2 +  z_6^2 = 1.
\end{array}
$$

The dual Hamiltonian is encoded in the variable $z_2$ and
corresponding Mukai-Sklyanin Poisson algebra gives the ``dual" Hamilton system
$$
\begin{array}{rcl}
{\dot z_1}=\{z_1,z_2\}= -z_3z_4z_5z_6\\
{\dot z_2}=\{z_2,z_2\} \equiv 0\\
{\dot z_3}=\{z_3,z_2\}= -z_1z_4z_5z_6\\
{\dot z_4}=\{z_4,z_2\}= -g^2z_1z_3z_5z_6\\
{\dot z_5}=\{z_5,z_2\}= -z_1z_3z_4z_6\\
{\dot z_6}=\{z_6,z_2\}=0
\end{array}
$$
which is clearly equivalent (``self-dual") to our initial
double-elliptic system (\ref{dellpol}).

{\bf Remark} It is worth noting that the dual Hamiltonian
(\ref{hamdual}) in the initial (``canonical") coordinates after the
Fourier rotation $p\to -q,\ q\to p$ goes to the Hamiltonian
(\ref{freeham}) of the ``free" ($g=0$) model corresponding to an
Arnold-Euler-Nahm $SU(2)$  top. Below we will show that there is a
transformation of the ``non-free" hamiltonian system with general
double-elliptic Hamiltonian (\ref{ham}) into a system describing a
pair of identical $SU(2)$-Arnold-Euler-Nahm tops.

\subsection{Algebro-geometric remarks}

The system of four quadrics (\ref{quads1})  can be considered as an
affine part, $S_s$, of the intersection surface $\bar{S_s} \subset {\CC}P^{6}$:
$$
\begin{array}{rcl}
\label{surface}
 x_1^2 - x_2^2 = x_0^2\\
 x_1^2 - x_3^2 =k^2 x_0^2 \\
 -g^2x_1^2 + x_4^2 + x_5^2=x_0^2 \\
 -g^2  x_1^2 +  x_4^2 + \tilde {k}^{-2} x_6^2=\tilde {k}^{-2}x_0^2.
\end{array}
$$
This is a singular complex surface (i.e. a two-dimensional algebraic
variety) which depends on the vector $\bar s =(1,k^2,g^2,{\tilde  k}^{-2})$.
A straight forward computation shows that (up to a
non-zero factor) the Jacobian for the above system is given
by the following matrix:
$$
\left(
\begin{array}{ccccccc}
x_1& -x_2& 0& 0& 0& 0& -x_0\\
x_1&  0&  -x_3& 0& 0& 0& -k^2 x_0\\
-g^2 x_1& 0& 0& x_4& x_5& 0& -x_0\\
0& 0& 0& 0& -x_5& {\tilde k}^{-2} x_6& ({\tilde k}^{-2} - 1)x_0
\end{array}
\right).
$$
There are therefore the two  following sets of singular points of
$\bar{S_s}$:
$$
1)\qquad x_0 = x_5 = x_6 = 0;\ x_4^2 = g^2x_1^2;\ x_1^2=x_2^2=x_3^2,
$$
which contains 8 points and
$$
2)\qquad x_0 = x_1 = x_2 = x_3 = 0;\ x_4^2 + x_5^2 = 0;\ x_4^2 + {\tilde
  k}^{-2}x_6^2 = 0,
$$
containing another 4 points.

We will argue below that our intersection surface belongs to the class of so-called
Adler-van Moerbeke surfaces in ${\CC}P^6$ which project to a (singular) Kummer
surface, which in its turn gives a Del Pezzo (elliptically fibered rational) surface
with 12 singular elliptic fibers (``1/2K3" surface). Our philosophy here is to
``embed" the 1-dimensional integrable system, associated with a complex curve (in our
case this is the ``base" elliptic curve of the coordinates $x_1$, $x_2$, $x_3$) into a
system associated with a 2-dimensional complex variety along with the deformation of
Hitchin systems proposed in \cite{Donagi}.

Our model has evidently only one degree of freedom so we are
interested in a section of the surface given by the ``energy level"
$H = z$, which is a hyperplane section in our case.
The canonical divisor $K_S \equiv (4\times 2 - 4 - 3)H = H$ is given by
a hyperplane section $H$ of the surface $\bar{S_s}$.

We will be interested in the hyperplane section ${\tilde {\cal C}}_z$ of the surface
by $x_5 = z x_0$, given by a fixed (``generic") complex number $z:\
{\tilde {\cal C}}_z
=\bar{S_s}\bigcap\{x_5= z x_0\}\subset {\CC}P^5$.

Let us compute the genus of this (singular) curve taking into account
that it passes through the 8 of the twelve singularities enlisted
above. The canonical divisor is calculated as
$K_{\tilde {\cal C}_z} = (8-6)H = {\cal O}_{\tilde {\cal C}_z}(2)$ and 
$\deg {\tilde {\cal C}_z} = (({\tilde {\cal C}_z}^2) +({\tilde {\cal C}_z},K_{S}) =  2^5$.
We have by the adjunction formula
$$
g({\tilde {\cal C}_z}) = 1/2(({\tilde {\cal C}_z}^2) +({\tilde {\cal C}_z},K_{S})) + 1 - \delta = 2^4
+ 1 - 8 = 9
$$

and we have that the genus of the (normalised) curve is 9.

Let consider the corresponding ("projected" )family of curves in ${\CC}P^4$ because
the last equation of the quadrics is a
superficial one due to the consistency condition between $x_5$ and $x_6$
(see 2.3).

The family ${\tilde C}_z$ represents a singular family of plane curves
$$
{\tilde C}_z\  : \ y^2 = x^2(x^2 -1)(x^2 -k^2)(g^2x^2 - (z^2 -1)),
$$
with $x =x_1$, $y = x_1x_2x_3x_4$.
This family is incomplete and has a double point at the origin for each
``generic" value of $z$. A completion of the ``generic" curves in the family
${\tilde C}_z$ may be achieved in two different ways which we will now
briefly discuss.

The first is by adding a point at infinity which is a completion of the affine
curve in $\CC^2$ to a singular curve $C_z \subset {\CC}P^2$ of genus
$g(C_z) = 2$. We have
$$
C_z\  : \ {\tilde Y}^2 {\tilde Z}^6 = ({\tilde X}^2 - {\tilde Z}^2)( {\tilde
X}^2- k^2
{\tilde Z}^2)(g^2{\tilde X}^2 - (z^2 -1){\tilde Z}^2),
$$
with $({\tilde X}:{\tilde Y}:{\tilde Z})$  homogeneous coordinates on
${\CC}P^2$ such that
$x = {\frac{\tilde X}{\tilde Z}}, y = \frac{\tilde Y}{\tilde Z}$.

The curve $C_z$ is a double covering of ${\CC}P^1$, $\pi_C :C_z \to {\CC}P^1$,
branching at 6 points. Its normalization $ E_{z}$ has genus 1 because the
family $C_{z}$ is a double cover of a family of elliptic curves
(an ``elliptic pencil''):
$$
E_z\ : Y^2 = (1 - X)(1 - k^2X)(g^2 -(z^2 -1)X),
$$
which is in turn a double cover of ${\CC}P^1$, $\pi_E :E_z \to {\CC}P^1$
with the invariants
$$
g_2(z) = 1/12\{ (k^4 - k^2 +1)(z^2 -1)^2  - k^2 g^2 (k^2 + 1)(z^2 -1) +
k^4 g^4\}\in {\cal O}_{z}(4);
$$
$$
g_3(z) = 1/48 k^2(z^2 -1)[(z^2 -1) + (k^2 +1)g^2][(k^2 +1)(z^2-1) +k^2
g^2] -
$$
$$
-1/216 [(k^2+1)(z^2-1)  + k^2 g^2]^3 -g^2/16 k^4(z^2-1)^2 \in
{\cal O}_{z}(6).
$$

For the second we can consider the complete curve $\bar C_z$ of genus 3
which is a normalization of the fiber product
$$
C_z \times_{{\CC}P^1}E_z :=\{(p,q)\in C_z \times E_z|\pi_C(p)=\pi_E(q) \}.
$$
The curve $\bar C_z$ is an ``\'etale double covering" of $C_z$.

Let us discuss the requirement for the intersection surface $S ({\bar S}_s)$
to be an abelian surface. In what follows we will suppose that it is.
The curve $C_z$ is irreducible of genus 2 curve with one double point admitting
an elliptic involution. It lies on the surface $S$. We describe the beautiful
geometric relations between the triple of the families of curves
${\bar C}_z, C_z$ and $E_z$ and the surface $S$ which is based on
\cite{barth}.

Fix the double covering $\pi : {\bar C}_z \to E_z$ and let $a_0$ be a
branch point.  Then we have the natural mapping
${\bar C}_z \to J({\bar C}_z):= Pic^0({\bar C}_z)$ by
$p\in {\bar C}_z \mapsto {\cal O}_{{\bar C}_z}(p-a_0)$.
If $\pi(a_0)= q_0$ then the elliptic curve $E_z$ is identified with
$Pic^0(E_z)$ by the mapping $q\in E_z \mapsto {\cal O}_{E_z}(q-q_0)$ so we
have the induced mapping $\pi^*: E_z \to J({\bar C}_z)$ such that
$\ker\pi^* = 0$. The compatibility of the involutions shows that the images
of $Pic^0(E_z)$ and $Pic^0(C_z)$ in $Pic^0(\bar C_z)$ are complementary.
The quotient of $Pic^0(\bar C_z)\cong J(\bar C_z)$ by the subgroup
$\ker(Pic^0(E_z)\to Pic^0(\bar C_z))$ is an abelian surface which is called
the Prym variety of the involution on $\bar C_z$ and denoted by
$Prym ({{\bar C}_z}/E_z)$. Then the surface $S$ is identified with
the quotient $J({\bar C}_z)/\pi^*(E_z)$ and is dual to
$Prym ({{\bar C}_z}/E_z)$ by the ``duality theorem" (th.1.12 in \cite{barth}).
Moreover, both surfaces $S$ and $Prym({\bar C_z }/E_z)$ are
(1,2)- polarized and the following diagram takes place:
$$
\xymatrix{&&0\ar[d]&\\
 &&Prym({\bar C}_z/E_z)\ar[d]\ar[dr]^{\Longleftrightarrow}&\\
 0\ar[r]&E_z\ar[r]^{{\bar \pi}^*}&J({\bar C}_z)\ar[d]\ar[r]& S\ar[r]&0\\
 &{\bar C}_z \ar[ur]\ar[r]^{{\bar \pi}}&E_z \ar[d]\ar[dr]^{\pi_{E}}& C_z
\ar[l]^{\pi}\ar[d]_{\pi_C}\ar[r]& J(C_z)\ar[ul]^{\simeq}\\
 &&0&{\CC}P^1&&
           }
$$

Here the symbol $\simeq$ means that the surfaces $Prym({\bar C_z}/E_z)$
and $J(C_z)$ are isogenous. The fact that $Prym({\bar C}_z/E_z)\simeq J(C_z)$ is
rather standard (in the framework of integrable systems we learnt it from the
beautiful book \cite{audin}).
The symbol $\Longleftrightarrow$ above means the duality between the
intersection surface $S$ and the Prym variety $Prym({\bar C}_z /E_z)$,
namely $S^{\vee}\Longleftrightarrow Prym({\bar C}_z /E_z)$ and
$(Prym({\bar C}_z /E_z))^{\vee}= S$ with the duality of (1,2)-polarizations.

On other hand, it is well-known (see \cite{LB}) that the Jacobian of a genus 2
curve of "type I" (which is here the case for $C_z$ which has affine equation
$y^2 = (x^2 -1)(x^2 - k^2)(g^2 x^2 - (z^2 -1))$) is isogenous to a product of
two elliptic curves. Hence the dual of the intersection
surface $S$ is isogenous to a product of two elliptic curves.
Naively this isogeny can be interpreted as the combination of the elementary
duality transformation of \cite{bmmm} and the Fourier-Legendre rotation applied
to the surface $S$ given by the quadric intersection.

 {\bf Remark} The ``non-generic" values of $z$ ($z^2 = 1,g^2 + 1, g^2 k^2 + 1)$
reduce the family ${\tilde C}_z$ to different examples of plane nodal rational
curves.

\subsection{Comparison with Beauville-Mukai systems}

When we fix the consistency condition
\beq
\tilde{k}^2 (1-E^2)=1-c^2
\eeq
a system of the following form emerges:
$$
\begin{array}{rcl}
 x_1^2 - x_2^2 =1 \\
 x_1^2 - x_3^2 =k^2 \\
-g^2 x_1^2 + x_4^2  +x_5^2  = 1.
\end{array}
$$

We would like to associate the intersection surface $S({\bar S}_s)$
with a double covering $\pi: S \mapsto {\CC}P^2$ € {\it a la}
Beauville-Takasaki (\cite{Beau, Taka}).

Let us suppose (according to the discussion in \cite{fgnr}) that the value
of the Hamiltonian $x_5 = z$ is a coordinate function on the open set
$U_0 = {\CC}P^1\setminus{\infty}$ and take homogeneous coordinates
$(X_0:X_1:X_2)$ on ${\CC}P^2$. Consider the surface
$S_0\subset {\CC}P^2\times {\AA}^1$ (${\AA}^1$ is an affine part of
${\CC}P^1$) given by the equation:
$$
X_0X_2^2 = X_1^3 + f(z)X_0^2X_1 + g(z)X_0^3,
$$
where $f(z)$ and $g(z)$ are polynomials. Again let us put
$y = x_1x_2x_3x_4$ and obtain the equation
\begin{eqnarray}
y^2 = x_1^2(x_1^2 - 1)(x_1^2 - k^2 )(g^2 x_1^2 -(z^2 -1)).
\end{eqnarray}
from our system of quadrics.
With help of the evident linear change
$$
X = X - 1/3(\frac{g^2}{z^2 -1} -\frac{1+k^2}{k^2})
$$
we can replace the curves of the cubic family
$$
Y^2 = (1-X)(1-k^2X)(g^2 - (z^2-1)X).
$$
by an equation of Weierstrass form. If we
consider the coordinates $Y,X$ as non-homogeneous for $(X_0:X_1:X_2)$ (ie
$Y= X_2/X_0$, $X= X_1/X_0$) the equation acquires the Weierstrass form:
\begin{eqnarray}
Y^2= X^3 + f(z)X + g(z),
\end{eqnarray}
where $f(z) \in {\cal O}(4), g(z) \in {\cal O}(6)$.

Let us produce the transition to the chart $U_1 = {\CC}P^1\setminus{0}$
with coordinate $1/z$:
$$
{\tilde z} = 1/z, {\tilde X} = X/z^2, {\tilde Y} = Y/z^3.
$$
We have
\begin{eqnarray}
{\tilde Y}^2 =  {\tilde X}^3 +  {\tilde f}( {\tilde z}) {\tilde X} +  {\tilde
g}( {\tilde z}),
\end{eqnarray}
where  ${\tilde f}( {\tilde z}) = f(z) z^{-4}, {\tilde g}({\tilde z}) =
g(z)z^{-6}$.
We can consider this as a surface $S_1\subset {\CC}P^2\times {\AA}^1$
(where ${\AA}^1 = U_1$) and hence the union $S_0\bigcup S_1$ correctly
defines a surface $f: S\in {\CC}P^1$ which is elliptically fibered.
In general it is a singular surface (at the points where the $J-$invariant
$J(z) = f(z)^3/\Delta(z)$, with $\Delta = 4f(z)^3 + 27g(z)^2 \in {\cal
O}(12)$ vanishes).

Therefore, we can suppose that the family of curves $\bar C_z$ covers
the elliptic pencil lying in (an open part of) a rational elliptic (``1/2 K3") surface $S$.
The rational elliptic fibrations (del Pezzo surfaces) can (roughly speaking)
be considered as a Kummer surfaces ${\cal K}$ modulo ${\ZZ}_2$.
The Kummer surface in our case is exactly the 2:1 projection of the
intersection (\ref{surface}):
$$
\begin{array}{cccc}
{\tilde {\cal C}_z}& \subset {\bar S}_s& \subset& {\CC}P^6\\
\downarrow& \downarrow && \downarrow\\
C_z&\subset {\cal K} &\subset&  {\CC}P^5.\\
\downarrow& \downarrow &\nearrow &\\
E_z& \subset {\cal K}/{\ZZ}_2
\end{array}
$$

{\bf Remark} This description fits well to the results of \cite{Sz}
where the full description of Kummer surfaces in ${\CC}P^5$ arising as
projections of the Adler-van Moerbeke \cite{AvM} family of abelian 
surfaces in  ${\CC}P^6$. Our description is corresponded to the first
type of the surfaces : a Kummer surface with 12 nodes. The explicit
quadratic equations of all 9 types of singular Kummer surfaces with
1:1 correspondence to the Adler-van Moerbeke surfaces can be found in
\cite{Sz}.

Let us give another geometric picture of the intersection surface.
In ``physical" language we are dealing with a family of ``supersymmetric
cycles" $(C_z,L)$ in $S$, where $C_z ^2 := [C_z]. [C_z] = 2g(C_z) - 2 = 2$
(because the genus $g(C_z) = 2$),
$[C_z]$ its homology class in $H_2(S,\ZZ)$
and $L$ is a flat $U(1)$-bundle ($:=$ a choice of a point on
$J(C_z)$).
We consider the linear system $|C_z|:=
{\PP}(H^0(S,{\cal O}(C_z)) = {\CC}P^2$ of all such curves.

The family $C_z$ is singular and we consider a compactification ${\bar J}(C_z)$
of the Jacobian variety $J(C_z)$ of $C_z$. This compactified Jacobian is a
fibration
$$
{\bar J}(C_z) \to J(E_z)
$$
with singular rational fiber. Here $J(E_z)$ is the Jacobian of the
normalisation such that the generalised Jacobian is the extension
$$
0 \to {\CC}^* \to J(C_z) \to J(E_z) \to 0.
$$
The union $\bigcup {\bar J}(C_z)$ has dimension $2g(C_z) =4$ and is a
fibration over $|C_z| = {\CC}P^2$.

In our case ${\cal O}(C_z)$ generates $Pic(S)$ so the family of the compactified
Jacobians carries a hyperkahler structure, moreover it is birationally
symplectomorphic to $Hilb^{[2]}(S)$ - the Hilbert scheme of 2 points on the
surface $S$ (see \cite{Beau, YZ, gnr}). This is a framework of
the so-called Beauville-Mukai integrable systems. Namely we have a
holomorphic Lagrangian fibration $\bigcup {\bar J}(C_z) \to |C_z|$ and an open
embedding of $Symm^2(S)\to \bigcup {\bar J}(C_z)$
(choosing a base point $p_{\infty}$ on $C_z$ we can use the
Abel-Jacobi map $AbJ :Symm^2(C_z) \to J(C_z)$,
$ AbJ(p_1,p_2) = {\cal O}_{C_z}(p_1 + p_2 -2p_{\infty} )$),
such that the following commutative diagram takes place
$$
\xymatrix{ &Symm^2(S)\ar[dr] \ar[r]& \bigcup {\bar J}(C_z)\ar[d]\\
& & |C_z|\\}.
$$

Hamiltonian in our case implies that we have a degree 6 polynomial defining a
hyperelliptic curve of genus 2:
$$
Y^2 = (cz^2 + u_1z + u_2)^3 + f(z)( cz^2 + u_1z + u_2) + g(z).
$$
Here we should consider the constant $c$ as a Casimir following
Takasaki's prescription in \cite{Taka}.
Various expressions of the canonical form are considered in
{Appendix B}.

\subsection{Comparison of two descriptions of the double-elliptic
  system}
Let us now compare the above description of the double-elliptic system with that
suggested in \cite{fgnr}. The two dimensional manifold which provides the
phase space in that paper was identified with the double covering of an
elliptically fibered two-dimensional manifold (K3 surface). The suggested
Hamiltonian was the coordinate on the base of the fibration while the
Hamiltonian of the dual system was identified with the
coordinate of the fiber.  It was supposed in \cite{fgnr} that
generically the K3 surface contained an
incomplete hyperelliptic genus five curve, and that is why Takasaki \cite{Taka}
in order to obtain an embedding of the one degree of freedom double-elliptic
system in to the scheme of Beauville had to set
four of his five parameters $u_i, i=1,\ldots,5$ to be equal
zero. But, as was shown  above, the manifold of \cite{bmmm} has the
structure of a double covering of an elliptically fibered rational surface.
Moreover in these terms the Hamiltonian of the  system is the coordinate on
the fiber, while the Hamiltonian for the dual system is the coordinate on
the base. Therefore we could consider the same argument as for the
double-elliptic system from \cite{fgnr} in the case of the rational
elliptic fibrations, (incomplete) hyperelliptic curves of
genus $2$ (a typical fiber of the dual system) and the double cover looks like a
hyperelliptic Jacobian of the genus 2 curve. This gives us a direct
embedding of our picture in the Beauville-Mukai systems {\it a la}
 Takasaki \cite{Taka},
taking the Hilbert scheme $S^{[2]}$ (or rather $Symm^2 (S)$) with the
symplectic form ${{dx^{(1)}\wedge dz^{(1)}}\over{y^{(1)}}} +{{dx^{(2)}\wedge
    dz^{(2)}}\over{y^{(2)}}}$ and the Abel -Jacobi map
$$
AbJ(p_1,p_2) = \left(\int^{p_1}_{p_{\infty}}{\frac{zdz}{y}} +
  \int^{p_2}_{p_{\infty}}{\frac{zdz}{y}};
 \int^{p_1}_{p_{\infty}}{\frac{dz}{y}} +
 \int^{p_2}_{p_{\infty}}{\frac{dz}{y}} \right) = (\phi_1,\phi_2)
$$
and
$$
AbJ^{*}\left({{dx^{(1)}\wedge dz^{(1)}}\over{y^{(1)}}} +{{dx^{(2)}\wedge
    dz^{(2)}}\over{y^{(2)}}}\right) = du_1\wedge d\phi_1 +  du_2\wedge d\phi_2.
$$
Here we suppose that $p_i = (z^{(i)},y^{(i)})\in S, i=1,2$
 We can
illustrate the situation by the following diagram of double-elliptic system  with the
interchange of the direct and dual systems.

$$
\begin{array}{ccccccccccc}
&S&\simeq &J(C_x)&\Longleftrightarrow &J(C_z)& \simeq &S^{\vee}&\\
&\swarrow^{q\in E_x}& &\searrow^{\varphi^D \in C_x} &&\swarrow_{\varphi \in
C_z}& &\searrow^{Q\in E_z = J(E_x)} \\
&x\in {\CC}P^1& &x\in {\CC}P^1 & &z\in {\CC}P^1& &z\in {\CC}P^1\\
\end{array}
$$

Moreover, now we are able to explain the ``naive"
definition of duality \cite{bmmm} arising from the anticanonical condition:
$$
dP\wedge dQ = - dp\wedge dq.
$$
Consider two ``dual" elliptic fibrations associated with the
double-elliptic system and with its dual in terms of the families of
projective Weierstrass cubics in ${\CC}P^2$:
$$
S\subset {\CC}P^2 : y^2 = z^3 + f(x)z + g(x), x\in {\CC}P^1,
$$
and
$$
{S^{\vee}}\subset {\CC}P^2 : y^2 = x^3 + {\tilde f}(z)x + {\tilde g}(z), z\in
{\CC}P^1.
$$
Here $z$, $y$ are nonhomogeneous coordinates in ${\CC}P^2$, with
canonical symplectic form $\Omega = {{dz\wedge dx}\over{y}}$ and similarly
$x$, $y$ are nonhomogeneous coordinates in ${\CC}P^2$, with the canonical
symplectic form $\Omega = {{dx\wedge dz}\over{y}}$.

Now the local action coordinate $I(x)$ is computed \cite{fgnr} as
$$
dI(x) = T(x)dx = \left(1/2\pi \oint_{A_x}dz/y\right)dx,
$$
where $[A_x]\in H_1({\mathcal E}_x, {\mathbb Z})$ is a chosen
$A$-cycle. Analogously, the dual action coordinate $I^{D}(z)$ satisfies
$$
dI^{D}(z) = T^{D}(z)dz = \left(1/2\pi \oint_{L_z}dx/y\right)dz,
$$
where $[L_z]\in H_1({\mathcal C}_z, {\mathbb Z})$.

Now we have the following chain of transformations:

$$
\Omega ={{dz\wedge dx}\over{y}}= - {{dx\wedge dz}\over{y}},
$$
$$
{{dz}\over{y}}\wedge {{dI}\over{T(x)}} = - {{dx}\over{y}}\wedge
{{dI^{D}}\over{T^{D}(z)}},
$$
or
$$
{{dz}\over{yT(x) }}\wedge dI = - {{dx}\over{yT^{D}(z) }}\wedge dI^{D}.
$$
Using the expressions for the angle variables
$d\varphi ={{dz}\over{yT(x) }}$ and
$d{\varphi}^D = {{dx}\over{yT^{D}(z) }}$ we obtain
$$
d\varphi \wedge dI = - d{\varphi}^D\wedge dI^{D}.
$$
Keeping in mind the interpretation of \cite{bmmm}, the equality above
amounts to
$$
K(k) dp^{Jac}\wedge dI = - K(\tilde k) dp^{{Jac}^{\vee}}\wedge dI^{D},
$$
or
$$
dP\wedge dQ = - dp\wedge dq,
$$
the starting point of \cite{bmmm}.

{\bf Remark} The duality transformation $x \to z$, $z\to x$ between two
elliptic pencils can  naively be interpreted as a local ``mirror
transformation" between two ``1/2K3" surfaces \cite{minah}: thinking
of $(z,x)$ of the first cubic as the Kahler and complex parameters we,
obtain that $(x,z)$ on the second becomes the corresponding pair of parameters.
Hence the duality at hand changes the ``Kahler" and ``complex" structures
of the cubic surfaces.

\subsection{Two coupling constants}

There exist natural generalizations of the double-elliptic system so far
described involving additional parameters expressing noncommutativity.
In terms of the intersection of quadrics the number of the
independent coupling constants amounts from the
number of  their independent coefficients. One simple case
involves such counting for the noncommutative $T^4$. In this case
all  couplings follow from noncommutativity, hence the maximal number of
parameters can be counted from the antisymmetric matrix $\theta_{ij}$
defining the Poisson bracket
\beq
\{x_i,x_j\}= \theta_{ij}
\eeq
and equals to six \cite{Ganor}. In the stringy framework
the couplings correspond to the  fluxes of rank-two field
switched on
along different directions in four-dimensional manifold.

One interesting example of two coupling
constants comes from the following system of quadrics
$$
\begin{array}{rcl}
 \tilde{g}^2x_4^2 + x_1^2 - x_2^2 =1 \\
 \tilde{g}^2x_4^2 + x_1^2 - x_3^2 =k^2 \\
 -g^2x_1^2 + x_4^2 + x_5^2=1 \\
 -g^2 x_1^2 +  x_4^2 +  x_6^2=\tilde {k}^{-2}
\end{array}
$$
which is written in the most symmetric form. This system yields
the following set of the Poisson brackets
$$
\begin{array}{c}
 \{x_1,x_2\}= \tilde{g}^2 x_4 x_5 x_3 x_6, \qquad
 \{x_1,x_3\}= \tilde{g}^2 x_4 x_5 x_2 x_6, \qquad
 \{x_2,x_3\}=0 \\
 \{x_5,x_1\}= -x_2 x_3 x_4 x_6, \qquad
 \{x_5,x_2\}= -x_1 x_3 x_4 x_6, \qquad
 \{x_5,x_3\}= -x_1 x_2 x_4 x_6 \\
 \{x_5,x_4\}= -g^2 x_1 x_2 x_3 x_6, \qquad
 \{x_5,x_6\}=0, \qquad
 \{x_6,x_4\}= -g^2 x_1 x_2 x_3 x_5.
\end{array}
$$

Now if $x_5$ is taken as the Hamiltonian then the coupling $g$ corresponds to
the ``interaction" of the coordinates, while $\tilde{g}$ corresponds to
the ``interaction" of momenta. The duality now acquires a more symmetric form
and correspond just to the  simultaneous interchange of $x_2$ and $x_5$ and
the  couplings. One could also consider examples of self-dual systems  by
choosing, say, $x_2x_5$ or $x_2+x_5$ as Hamiltonians.

\section{Degenerations}

Let us briefly consider possible degenerations and limits of the
model. If one sends $\tilde k$ to zero the dynamical system
degenerates to the elliptic Ruijsenaars model and the corresponding four
dimensional manifold reduces to $[T^2 \times {\CC}^*]_{g}$ which will be
explicitly described below. At the next step one can reduce the
model to the elliptic Calogero model with the manifold $[T^2 \times {\CC}]_{g}$.
Alternately, if $k$ is sent to zero the system reduces to the one which
is dual to the elliptic Ruijsenaars and Calogero models respectively.

The constructions involving quadrics provide a very explicit description of the
noncommutative manifolds above. Indeed let us consider limit $\tilde k \to 0$
which amounts to the system of three quadrics in ${\CC}^5$
$$
\begin{array}{rcl}
 x_1^2 - x_2^2 =1 \\
 x_1^2 - x_3^2 =k^2 \\
-g^2 x_1^2 + x_4^2  +x_5^2  = 1.
\end{array}
$$
Geometrically this looks like a cylinder which fibered over the elliptic curve.
The corresponding deformed four-manifold for the Calogero system $[T^2 \times
{\CC}]_{g}$ is  a ``pinched" ${\CC}P^1$ fibering over the elliptic curve.

Let us discuss the relation with the Sklyanin algebra
\cite{sklyanin}. To this end
let us interpret the Poisson bracket relations in the Sklyanin
algebra
\beq
\{S_\alpha ,S_0\} = 2 (J_{\beta}-J_\gamma)\,S_{\beta}
S_{\gamma}
\eeq
\beq
\{S_\alpha ,S_ \beta\} = 2 S_0 S_\gamma
\eeq
as an example of the Mukai-Sklyanin algebra generated by two quadrics
in ${\CC}P^4$. The quadrics
\begin{eqnarray}
&2 Q_1=\sum_{n=1}^{3} S_{n}^2 \\
&2 Q_2=S_{0}^2 +\sum_{n=1}^{3} J_n S_{n}^2
\end{eqnarray}
coincide with the center of the Poisson bracket algebra. Hence the
Sklyanin algebra fits into the general scheme.

Now recall the observation made in \cite{kriza} that the Hamiltonian of the
elliptic Ruijsenaars system coincides with the generator of the Sklyanin algebra
$S_0$. On the other hand, following our approach, the double-elliptic
Hamiltonian coincides with the coordinate $x_5$.
This means that in the limit of vanishing $\tilde {k}$ the system of four
quadrics in ${\CC}P^6$ reduces to a system of two quadrics
in ${\CC}P^3$ corresponding to the Sklyanin algebra.

The degeneration to the trigonometric
Ruijsenaars model can be performed in a similar way and the corresponding
Hamiltonian can be expressed in terms of the coordinates which follow from the
realisation of $U_{q}(SL(2))$ in terms of the intersection of two quadrics in
${\CC}^4$. The explicit degeneration in this direction goes as follows.
Let us first consider the algebraic structure intermediate between the
Sklyanin algebra and $U_{q}(SL(2))$.
Following \cite{gz}, we define generators $A,\, B, \, C, \, D$
of the degenerate Sklyanin algebra  at "small" $h=e^{i\pi \tau} (\Im
\tau \to 0)$:

\begin{eqnarray}
&&
A=-\, \frac{h^{-1/2}}{2\sin \, 2\pi \eta }
(\cos \,\pi \eta \, S_0 +i\sin \, \pi \eta \, S_3 ),
\nonumber\\
&&
D=-\, \frac{h^{-1/2}}{2\sin \, 2\pi \eta }
(\cos \,\pi \eta \, S_0 -i\sin \, \pi \eta \, S_3 ),
\nonumber\\
&&
C=-\, \frac{h^{1/2}}{2\sin \, 2\pi \eta }
(S_1 -iS_2 ),
\nonumber\\
&&
B=-\, \frac{h^{-3/2}}{8\sin \, 2\pi \eta }
(S_1 +iS_2 ).
\label{t15}
\end{eqnarray}

The generators $A,B,C,D$ satisfy the quadratic
algebra \cite{gz}:
\begin{eqnarray}
&& DC=e^{2\pi i\eta }CD,\;\;\;\;\;\;CA=e^{2\pi i\eta }AC,
\nonumber \\
&& AD-DA=-2i\,\sin ^{3} 2\pi \eta \, C^2, \nonumber \\
&&BC-CB=\frac{A^2 -D^2 }{2i\sin \, 2\pi \eta }, \nonumber \\
&& AB-e^{2\pi i\eta }BA=e^{2\pi i\eta }DB-BD=
\frac{i}{2}\sin \, 4\pi \eta \,(CA-DC).
\label{algebra}
\end{eqnarray}

The Casimir elements are the following quadrics
\begin{eqnarray}
&&Q_1 = e^{2\pi i \eta }AD -\sin ^{2}2\pi \eta \, C^2,
\nonumber \\
&&Q _2 = \frac{e^{-2\pi i\eta }A^2 +e^{2\pi i \eta }D^2 }
{4\sin ^{2}2\pi \eta }
-BC -\cos \, 2\pi \eta \, C^2 .
\label{casimirs}
\end{eqnarray}
and this algebra fits within the general formula
for a generalised Mukai-Sklyanin algebra.
Some generalisations of this picture for the trigonometric many-body system
can be found in \cite{hasegawa}. It would be extremely interesting to
present the elliptic Ruijsenaars-Schneider many-body systems
explicitly in a similar manner in terms of the
quadratic elliptic algebras of \cite{FO} or their tensor products.

To get the $U_{q}(SL(2))$ itself from the algebra above it is
necessary to consider the contraction $\epsilon \to 0$:
$$
B\to B,\ A\to \epsilon A,\  D\to \epsilon D,\ C\to \epsilon^2 C.
$$
and take the following quadrics given by the Casimirs
\begin{eqnarray}
&& Q_1 = AD
\nonumber \\
&& Q_2 =\frac{q^{-2}A^2 +q^{2}D^2 }{4(q-q^{-1})} - BC,
\end{eqnarray}
where $q = \exp{i\pi\eta}$.
The Poisson brackets following from the algebra of the quadrics
reproduce the standard Poisson structure of $U_{q}(SL(2))$.

One further remark concerns the degeneration to the Calogero model.
In principle we could obtain this by two paths. First, we can represent the
system in terms of the quadratic Hamiltonian with linear Poisson bracket.
To do this it is useful to consider the limit of large value
of the Casimir $Q_2$ and express
$S_0$ in terms of other generators of the Sklyanin algebra. Expanding the
root to first order we immediately get the representation
of the elliptic Calogero system as an elliptic rotator
\begin{equation}
H=\sum_{n=1}^3 J_i S_i^2
\end{equation}
which reduces to the standard form after a redefinition
of the coordinate. However we could also consider
the quadratic algebra using the degeneration of the
action of $x_0$ on the function $f(u)$
\begin{equation}
x_0 f(u) =\frac{\theta_{11}(\eta)(\theta_{11}(2u - l\eta)f(u +\eta) -
\theta_{11}(-2u - l)f(u -\eta))}{\theta_{11}(2u)}
\end{equation}
where $l$ is the spin of representation corresponding to the coupling
constant in the
dynamical system. To get the elliptic Calogero system we have
to send $\eta \rightarrow 0$ hence the difference operator
becomes the differential one.

For completeness let us note that the periodic Toda two-body system  can also
be described in terms of a top on the quadratic algebra. To this end
consider the following Poisson algebra
$$
\begin{array}{c}
 \{A_1,A_2\}=2A_3(4-A_2),\qquad
 \{A_3,A_2\}=A_2,\qquad
 \{A_1,A_3\}=A_1+A_3^2.
\end{array}
$$
Then the Hamiltonian of the Toda system is the linear function
\beq
H_{Toda}=1/2(A_1+A_2)
\eeq
This picture can also be considered as the further
degeneration from the elliptic Calogero model via the Inozemtzev limit.

\section{Other incarnations of the double-elliptic system}

\subsection{Double-elliptic systems and decoupled Nahm tops}

We are now able to give another polynomial description of the
double-elliptic system observing that it has the form of Fairlie's
 ``elegant" integrable system \cite{Fairlie}
for $n=4$. Ignoring the unimportant coordinate $x_6$ we have,
\begin{eqnarray}
\label{elegant}
& {\dot x_1} = x_2x_3x_4\\
& {\dot x_2} = x_1x_3x_4\\
& {\dot x_3} = x_1x_2x_4\\
& {\dot x_4} = g^2x_1x_2x_3.
\end{eqnarray}

This system admits a beautiful description as a {\it decoupled} system of
Euler-Nahm tops after the following change of variables:
$$
\begin{array}{rcl}
 u_+ = x_3x_4 + g x_1x_2, \qquad
 v_+ = x_2x_4 + g x_1x_3 , \qquad
 w_+ = x_1x_4 + g x_3x_2\\
 u_- = x_3x_4 - g x_1x_2, \qquad
 v_- = x_2x_4 - g x_1x_3 , \qquad
 w_- = x_1x_4 - g x_3x_2 .
\end{array}
$$
In the terms of the new variables the double-elliptic system (\ref{elegant}) is
equivalent to
\begin{eqnarray}
\label{Nahm_{+}}
& {\dot u_{+}} = v_{+}w_{+}\\
& {\dot v_{+}} = w_{+}u_{+}\\
& {\dot w_{+}} = u_{+}v_{+}.
\end{eqnarray}
and to
\begin{eqnarray}
\label{Nahm_{-}}
& {\dot u_{-}} = v_{-}w_{-}\\
& {\dot v_{-}} = w_{-}u_{-}\\
& {\dot w_{-}} = u_{-}v_{-}.
\end{eqnarray}
Geometrically this change of variables means a passage from the
intersection of the quadrics to the direct (``decoupled") product of two
elliptic curves ${\cal E_{+}}\times{\cal E_{-}}$ given by the Casimirs
of the models (\ref{Nahm_{+}}) and (\ref{Nahm_{-}})
$$
\begin{array}{rcl}
{\cal E_{+}}:
 u_{+}^2 - v_{+}^2 =k^2(E^2-1) \\
 u_{+}^2 - w_{+}^2 =(k^2 -2)(E^2 -1)
\end{array}
$$
and
$$
\begin{array}{rcl}
{\cal E_{-}}:
 u_{-}^2 - v_{-}^2 =k^2(E^2-1) \\
 u_{-}^2 - w_{-}^2 =(k^2-2)(E^2-1).
\end{array}
$$

This result is reminiscent of the result of of Ward \cite{ward} that the
second order differential operator with Lam\'e potential $n(n+1)/2 sn^2(q|k)$
can be factorized into a product of the first order matrix operators
$(\pa + A)(\pa - A) $ with the matrix $A = (A_1,A_2,A_3)$ satisfying the Nahm
equations
$$
{\dot A_i} = \epsilon_{ijk}[A_j,A_k].
$$
This property manifests a hidden supersymmetry underlying the quantum
mechanics with Lam\'e potential. The factorization of the double-elliptic
system into two decoupled tops
suggests  the existence of a hidden SUSY in this case too.

\subsection{The Double-elliptic system as an example of a Nambu-Hamilton system}

Our polynomial parametrization of the double-elliptic system also
provides and example of a Nambu-Poisson structure.
Such a structure was introduced by Nambu in 1973 \cite{nambu} as a
natural generalisation of the Poisson structure
(which corresponds below to $n=2$).
Such structures were recently extensively studied in \cite{tacht}.
Recall that a Nambu-Poisson manifold $M^n$ is is a manifold
endowed with an antisymmetric $n$-vector $\eta \in \Lambda^n(TM)$.
The field $\eta$ defines an $n$-ary operation:
$$
\bigl\{,...,\bigr\}: C^{\infty}(M)^{\otimes n}\mapsto
C^{\infty}(M)
$$
such that the three properties are valid:
\begin{enumerate}
\item antisymmetry:
$$
\bigl\{f_1,...,f_n\bigr\} =
(-1)^{\sigma}\bigl\{f_{\sigma(1)},...,f_{\sigma(n)}\bigr\},\qquad \sigma
\in S_n;
$$

\item coordinate-wise ``Leibnitz rule" for any $h\in C^{\infty}(M)$:
$$
\bigl\{f_1h,...,f_n\bigr\} = f_1\bigl\{h,...,f_n\bigr\} +
h\bigl\{f_1,...,f_n\bigr\};
$$

\item  The ``Fundamental Identity" (which replaces the Jacobi Identity):
$$
\bigl\{\bigl\{f_1,...,f_n\bigr\},f_{n+1},...,f_{2n-1}\bigr\} +
\bigl\{f_n,\bigl\{f_1,...,({f_n})^{\vee}f_{n+1}\bigr\},f_{n+2},...,f_{2n-1}
\bigr\}
+
$$
$$
+
\bigl\{f_n,...,f_{2n-2},\bigl\{f_1,...,f_{n-1},f_{2n-1}\bigr\}\bigr\}
= \bigl\{f_1,...,f_{n-1},\bigl\{f_{n},...,f_{2n-1}\bigr\}\bigr\}
$$
for any $f_{1},...,f_{2n-1}\in C^{\infty}(M)$.

\end{enumerate}

Dynamics on a Nambu-Poisson manifold is defined by $n-1$
Hamiltonians $H_1,...,H_{n-1}$ giving the Nambu-Hamiltonian system
$$
{{dx}\over{dt}} = \bigl\{H_1,H_2,...,H_{n-1},x\bigr\}.
$$
The most common example of a Nambu-Poisson structure is the
so-called ``canonical" Nambu-Poisson structure on ${\CC}^n$ with
coordinates $x_1,...,x_n$:
$$
\bigl\{f_{1},...,f_{n}\bigr\}= Jac(f_{1},...,f_{n})={{\partial
(f_1,...,f_n )}\over{\partial(x_1,...,x_n)}}
$$
Our constructions involving quadrics give non-trivial examples
of a Nambu-Hamilton dynamical system. Namely, the system of
three quadrics in ${\CC}^5$
$$
\begin{array}{rcl}
 x_1^2 - x_2^2 =1 ,\qquad
 x_1^2 - x_3^2 =k^2 ,\qquad
-g^2 x_1^2 + x_4^2  +x_5^2  = 1.
\end{array}
$$
admits a section by the choice of the level $x_5 = E$ and the
1-parameter intersection of three quadrics in ${\CC}^4$
$$
\begin{array}{rcl}
Q_1 = x_1^2 - x_2^2 =1 \\
Q_2 = x_1^2 - x_3^2 =k^2 \\
Q_3 = -g^2 x_1^2 + x_4^2 = 1 - E^2.
\end{array}
$$
define a Nambu-Hamilton system,  which is our double-elliptic system:
$$
{{dx_i}\over{dt}} = \bigl\{Q_1,Q_2,Q_3,x_i\bigr\}.
$$

\section{N-body systems}

In this section we discuss the challenging problem
of constructing the many-body generalisations of double-elliptic systems.
We will not present an explicit realisation of such systems,
but some problems and proposals concerning the matter will be given.
Let us emphasise that any finite dimensional two-body system can be generalised
in at least in two directions.

The first one involves spin chain type systems.
The starting point to be generalised is the Sklyanin Lax operator
expressed in terms of the generators of the Sklyanin algebra obeying  the
$RLL=LLR$ relations with Belavin's elliptic $R$-matrix. The corresponding
$N$-body system is the $XYZ$ spin chain. Now we would like to
construct the local Lax operator expressed in terms
of the generators of a generalised Mukai-Sklyanin algebra
of $n$ quadrics for $n>2$. It can be shown that it is impossible
to introduce  a local Lax operator whose matrix elements
are such generators assuming that Belavin's $R$ matrix intertwines it.
Therefore to pursue this line of generalisation one has to develop some
generalisation of the elliptic $R$-matrix.

Another line of generalisation would be to seek  the putative
many-body double-elliptic system as a generalised Hitchin system.
Hitherto the most complicated system treated in this way has been the elliptic
Ruijsenaars model which can be obtained via  the Hamiltonian reduction
procedure from the following moment map equation \cite{aru}
\beq
g(z)h(z)g^{-1}(z)h^{-1}(qz)=\mu
\eeq
where $g$ and $h$ are elements of affine $SL(N)$ and $\mu$
is the fixed level of the moment map.

To obtain the double-elliptic system one presumably has to solve the following
equation
\beq g(z)h(z)g^{-1}(\tilde{q}z)h^{-1}(qz)=\mu \eeq
which can be considered as the moment map equation for the Heisenberg double
of the double affine algebra. The symplectic manifold to be constructed is the
generalisation of the cotangent bundle to the double affine
group which provides the phase space for the elliptic Ruijsenaars model. If the
solution to this equation could be found (in some particular gauge)
the Hamiltonian of the many-body double-elliptic system would be given by
$\int Tr g(z)$. The results obtained in \cite{hasegawa}  are of help for
this approach.

To link the Heisenberg double approach with the geometrical considerations
of this paper it is instructive to adopt the fundamental group interpretation
of the moment map equations. The simplest equation of such type corresponding
to the trigonometric Ruijsenaars model looks like
\beq
ghg^{-1}h^{-1}=\mu.
\eeq
This has no coordinate dependence and just represents the relation
following from the fundamental group on the torus with one marked point.
To discuss the double-elliptic case in a similar manner we have to
consider connections on the whole four dimensional manifold.
Actually we are interested in a twisted bundle which in the simplest example
of the torus  results in t'Hooft's consistency condition for the twists
\cite{thooft}
\beq
\label{twists}
\Omega_i(z+a_i)\Omega_j(z+a_j)\Omega_i^{-1}(z+a_i)\Omega_j^{-1}(z+a_j)=Z_{ij}.
\eeq
Here $Z_{ij}$ belongs to the center of the group
and represents the matrix of fluxes. The torus case is too simple to generate
the double-elliptic system. However if we solve (\ref{twists})
then the hamiltonian $\int Tr \Omega_i$ would provide
a very degenerate example of such a ``double-elliptic '' system.

One more powerful approach is based on the separation of variables procedure.
Since we have identified the phase space of the two-body double-elliptic system
one could imagine that the Hilbert scheme of the $n$-points on this manifold
would solve the problem. However, a serious technical difficulty arises since
there is no effective description of Hilbert schemes of points on generic four
dimensional manifolds. A related issue involves the identification of
integrable many-body systems in terms of noncommutative instantons \cite{bn}.
To obtain the description of the double-elliptic system in terms of
noncommutative instantons the ADHM description of instantons on commutative
and noncommutative K3 manifolds should be developed. (See \cite{Ganor} for
steps in this direction.)

\section{The relation to D=6 SUSY gauge theories with
massive adjoint matter}

Let us briefly discuss the possible application of double-elliptic systems.
It is known that there is the map between classical integrable many-body
systems of Hitchin type and their generalisations
(\cite{BKr, gm} and the references therein) with supersymmetric gauge theories
in $d=4$, $5$, $6$ dimensions with the different matter content.
This relation can be briefly formulated as follows.
The  solution to the classical equation of
motion of the integrable system provides
the spectral curve embedded into
the higher dimensional compactification manifold.
The gauge theory is defined on the worldvolume
of the M5 brane wrapped around the spectral curve.
The particular integrable system
is in one to one correspondence with known  supersymmetric gauge theories.
Namely, the space of integrals of motion is mapped into the Coulomb branch
of the moduli space in the gauge theory, while the modular parameters of the
spectral curve of the integrable system amount to the running coupling
constants on the field theory side.

Previously the  elliptic Ruijsenaars integrable system, of Hitchin type, was
identified with an $N=1\ d=5$ SUSY gauge theory with one compact dimension
and adjoint hypermultiplet \cite{nikita5, ggm, bmmm2}.
The double-elliptic systems we are looking for presumably correspond to the
next step, namely an  $N=1$, $d=6$ gauge theory with adjoint hypermultiplet
when two dimensions are compactified.
There are two natural elliptic moduli in the gauge theory: one is the
complexified coupling constant while the second is the modulus of the
compact torus in the fifth and sixth dimensions. These are to
be identified with the moduli $k$, and $\tilde{k}$ in the quadrics above.
The coupling constant $g$ as before can be related to the mass of the adjoint
field. We suggest that the genus 2 curve (\ref{spectral}) we have found for the two-body
double-elliptic system plays the role of the Seiberg-Witten curve providing the
renormalised coupling constant in $d=6$ gauge theory with two compact
dimensions.

The Coulomb branch of the moduli space of this theory can be also described, via
an additional compactification of one further dimension, and consideration  of
the resulting $2+1$ theory with $N=4$ SUSY  as the theory of NS5 branes
compactified on ${\TT}^3$ \cite{Ganor}. The moduli space  of this theory
can be mapped into the moduli space of noncommutative instantons on ${\TT}^4$
and K3 . The noncommutativity corresponds to the fluxes of $B$ field along some
directions. Generically fluxes can be switched on in all directions,
corresponding to the complete set of twists in the language of
\cite{Ganor}. The system considered in this paper corresponds to  a flux only
along the fiber of the elliptic fibration which plays the role of the coupling,
constant in the dynamical system and the mass of adjoint in the field theory.
To link this with the integrable system one could adopt the description of the
instanton moduli space via the spectral curve endowed with the spectral bundle
\cite{freedman}. This description emerges after a
$T$ duality transformation of the initial four dimensional manifold along two
dimensions.

It has been argued that the instanton spectral curve coincides exactly with the
Seiberg-Witten curve describing the Coulomb branch of the moduli space of some
$N=2$ theory. The reason for this relation is that moduli space of the
instantons on noncommutative ${\TT}^4$ is equivalent to the moduli space of the
little string theory on ${\TT}^3$. The twists correspond to the structure of
the ${\TT}^2$ fibration over the the base ${\TT}^2$. In the
generic situation there are twists over the fiber and base tori. Geometrically
the situation  matches with the picture of the intersection of the quadrics
above. The coupling constants correspond to the twists and
it is therefore natural to conjecture that the double-elliptic $n$ body system
(or possibly a degenerated version) is related to $n$, $U(1)$ instantons on
noncommutative ${\TT}^4$.

\section{Conclusion}

In this paper we have elaborated in some detail the simplest double-elliptic
systems and their relation to the generalised Mukai-Sklyanin finite dimensional
algebras. The previously discovered Hamiltonians for these systems  were
placed in an algebro-geometric setting allowing various connections with
Nahm tops to be made.

The generalisation to the many-body case still remains a challenging problem.
We have however identified several mathematical problems to be solved to
make progress in this direction. Namely, it is highly desirable to find the
generalisation of the $R$-matrix related to the generic
algebra of quadrics, to develop the ADHM description of Hilbert scheme of
points on intersection of the quadrics and to study further the ADHM
description of instantons on noncommutative K3 manifolds.
It is also necessary to find the  ``$SU(N)$"
generalisation of the algebra of quadrics
which would involve a larger number of generators but keeping
the power of the generators involved (cubic, quartic e.t.c)
the same as for ``SU(2)'' case.
Such a generalisation is known  \cite{FO}
only for the Sklyanin algebras of 2 quadrics. The corresponding
quadratic Feigin-Odesskii algebras involve an arbitrary number
of generators and can be built from the vector bundles
on the elliptic curve.
The last problem to be mentioned
is that of finding a precise formulation of the notion of
the Heisenberg double to a double affine algebra. It would be also
very interesting to combine the ideas from this paper with the recent
two-dimensional generalisation of the Calogero and Hitchin type
systems suggested in \cite{Krich},\cite{LOZ}.

Besides the problems above, which can be considered as primarily
mathematical ones, there are a number of issues
relevant for the physics of D=6 gauge theories. First,
let us emphasise that the genus two spectral curve
(\ref{spectral}) provides the low energy
effective action and the spectrum of BPS particles
in D=6 theory. It would be interesting to pursue
further the duality arguments relating the theories
in different dimensions.
One more interesting
possibility along this way involves the idea of  the
deconstruction of
dimensions, starting from D=4 theory.
The deconstruction of D=6 theories with two
compact dimensions has been  considered
recently \cite{ah}. We expect that the double elliptic
system considered here could be  derived
from the ``double lattice'' of the integrable models
corresponding to D=4 if the nonperturbative
effects would be taken into account. This
correspondence has been proven for
the D=5 models with one
compact dimension which can be derived from the ``lattice''
provided by  the quiver type D=4 theory
\cite{csaki}.

\section{Acknowledgements}
We are grateful to M. Audin, A. Marshakov, A. Mironov, A. Morozov, N. Nekrasov, M.
Olshanetsky, I.Reider for the useful discussions. A substantial part of this work was
done during the visits of A.G. to Angers University under the Senior Researcher's
Grant of EGIDE to which he is grateful. H.W.B. is grateful to IHES and the Newton
Institute for support during which this work has progressed. The work was supported in
part by grants CRDF-RP1-2108 and INTAS-00-00334 (A.G) RFBR 99-01-01669, CRDF-RP1-2254
and INTAS-00-00055 (A.O), INTAS-99-1705, RFBR-01-01-00549 and by grant for support of
scientific schools 00-15-96557(V.R).

\section{Appendix}

We give for completeness the computations showing that the double-elliptic
Hamiltonian system in canonical variables is rewritten in the form
(\ref{quad}) under the change of variables  (3-8).

\subsection{A Hamiltonian equations}

We begin with the following relations
$$
\frac{\partial \sn(u|k)}{\partial k}=
\frac{\cn(u|k) \dn(u|k)}{k k'\sp2}\left[ -E(u)+k'\sp2 u +
k\sp2 \sn (u|k)\cd (u|k)\right]
$$
$$
\frac{\partial \cn(u|k)}{\partial k}=
\frac{\sn(u|k) \dn(u|k)}{k k'\sp2}\left[ E(u)-k'\sp2 u -k\sp2
\sn (u|k)\cd (u|k)\right]
$$
$$
\frac{\partial \dn(u|k)}{\partial k}=
\frac{k\sn(u|k) \cn(u|k)}{k'\sp2}\left[ E(u)-k'\sp2 u -
\sn (u|k)\dc (u|k)\right]
$$
where $k'\sp2=1-k\sp2$, with $\cd (u|k)=\frac{\cn(u|k)}{ \dn(u|k)}$
and so forth. Here $E(u)=E(u,k)$ is the elliptic integral of the second kind
$$
E(u)=\int_0\sp{\sn(u|k)}\sqrt{\frac{1-k\sp2 t\sp2}{1-t^2}}dt
=\int_0\sp{u}\dn\sp2(v|k)dv.
$$

Only the first of these relations needs to be established,
with the latter two following from partial differentiation of the
identities
$$
\sn\sp2(u|k)+\cn\sp2(u|k)=1
$$
and
$$
\dn\sp2(u|k)+k\sp2 \sn\sp2(u|k)=1
$$
respectively.

To establish the first we take the partial derivative with respect to
$k$ of the definition
$$u=\int_0\sp{\sn(u|k)} \frac{dt}{\sqrt{(1-t^2)(1-k\sp2 t\sp2)}}$$
giving
$$
0=\frac{1}{\cn(u|k)\dn(u|k)}\frac{\partial \sn(u|k)}{\partial k}
+ k\int_0\sp{\sn(u|k)} \frac{t\sp2 dt}{(1-k\sp2 t\sp2)
\sqrt{(1-t^2)(1-k\sp2 t\sp2)}}.
$$
Upon change of variable we have
$$
-\frac{1}{k\cn(u|k)\dn(u|k)}\frac{\partial \sn(u|k)}{\partial k}=
\int_0\sp{u}\frac{\sn\sp2(u|k)}{\dn\sp2(u|k)}du
$$
which is straightforwardly integrated to yield the first identity.

Let us set
$$
\alpha=\sqrt{1+\frac{g\sp2}{\sn\sp2(q|k)}},\qquad
\beta=\sqrt{1+\frac{g\sp2 \tilde k\sp2}{\sn\sp2(q|k)}},\qquad
\Lambda=\frac{\tilde k
\alpha}{\beta}.
$$
The equations of motion for the Hamiltonian
$H=\alpha(q) \cn(\beta(q) p|\Lambda)$ yields
$$\dot q=\frac{\partial H}{\partial p}=
-\alpha \beta \sn(\beta p|\Lambda) \dn(\beta p|\Lambda)
$$
and
$$
\dot p=-\frac{\partial H}{\partial q}=
-\partial_q \alpha \cn(\beta p|\Lambda)+p \alpha
\sn(\beta p|\Lambda) \dn(\beta p|\Lambda)\partial_q \beta
-\alpha \partial_\Lambda \cn(\beta p|\Lambda)\partial_q \Lambda.
$$
Here
$$
\partial_q \Lambda= -g\sp2\tilde k (1-\tilde k\sp2)\frac{\cn(q|k)
\dn(q|k)}{\sn\sp3(q|k)\alpha\beta\sp3}.
$$

We note that $\dot p$ is invariant under
$p\rightarrow p+\frac{2n K}{\beta}$ even though the explicit momentum terms
and the elliptic integral of the second kind in the last term individually are
not.

Now introduce the quadrics
$$
\begin{array}{rl}
x_1\sp2 -x_2\sp2&=1\\
x_1\sp2 -x_3\sp2&=k\sp2 \\
\alpha\sp2&=1+ g\sp2 x_1\sp2 \\
\beta\sp2&=1+ g\sp2\tilde k\sp2 x_1\sp2
\end{array}
$$
which may be realised by
$$x_1=\frac{1}{\sn(q|k)},\ \ \
x_2=\frac{\cn(q|k)}{\sn(q|k)},\ \ \
x_3=\frac{\dn(q|k)}{\sn(q|k)}.
$$
Consider further the quadrics
$$
\begin{array}{rl}
y_1\sp2 +y_2\sp2&=1\\
\Lambda\sp2 y_1\sp2 +y_3\sp2&=1
\end{array}
$$
realised by
$$
y_1=\sn(\beta p|\Lambda),\ \ \
y_2=\cn(\beta p|\Lambda),\ \ \
y_3=\dn(\beta p|\Lambda).
$$

Upon setting
$$
x_4=\alpha y_1,\ \ \
x_5=\alpha y_2,\ \ \
x_6=\beta y_3
$$
we  recover (\ref{quads1}) and find that $H=x_5$. Further
$$
\dot x_4 =\sn(\beta p|\Lambda)\partial_q \alpha \dot q+
\alpha \cn(\beta p|\Lambda)\dn(\beta p|\Lambda) \left(
p\partial_q \beta \dot q +\beta \dot p\right) +
\alpha
\partial_\Lambda \sn(\beta p|\Lambda)\partial_q\Lambda\dot q.
$$

We see
$$
\dot x_4=  g\sp2 x_1 x_2 x_3 x_6
$$
For this we don't actually need the explicit expressions for
$\partial_\Lambda \sn(\beta p|\Lambda)$ and $
\partial_\Lambda \cn(\beta p|\Lambda)$ for these appear in the combination
$$
-\alpha\sp2 \beta \partial_q\Lambda \dn(\beta p|\Lambda)
\left(\sn(\beta p|\Lambda)\partial_\Lambda \sn(\beta p|\Lambda)+
\cn(\beta p|\Lambda)\partial_\Lambda \cn(\beta p|\Lambda)\right)
$$
which vanishes as a result of the first identity given earlier.

\subsection{B Canonical 2-form in different parametrizations}

Let us compute the expressions for the canonical 2-form $dp\wedge dq$ given
in the local coordinates of the initial elliptic curves as a 2-form on the
phase surface in the different parametrizations.

We begin with the following identity in the region $x_5\neq 0$
$$
\Omega = dp\wedge dq = {\frac{dx_1\wedge dx_4}{x_2x_3x_5x_6}}.
$$
Indeed, we have from the obvious relations
$$
dp\wedge dq = \frac{dH\wedge dq}{\frac{\pa H}{\pa p}} = \frac{dx_5\wedge
dx_1}{x_2x_3x_4x_6}.
$$
and (\ref{dell1}),(\ref{alpha}) and (\ref{beta}) amount to
$dx_5\wedge dx_1 ={\frac{x_4}{x_5}}dx_1\wedge dx_4$.

In the notation of Section 2
$$
\Omega = \frac{dX\wedge dz}{x_6 Y}.
$$
This follows from the identities
$$
\Omega =\frac{dz\wedge dx_1}{x_2x_3x_4x_6} =\frac{dz\wedge
 dx}{x_1x_2x_3x_4x_6} =\frac{dz\wedge dx}{yx_6} = \frac{dz\wedge
 d(1/X)}{Y/X^2 x_6} = \frac{dX\wedge dz}{Yx_6}.
$$


\begin{thebibliography} {99}


\bibitem{kks}
M.~Olshanetsky, A.~Perelomov
Lett. Nuovo. Cim. ,{\bf 16} (1976) 333;
{\bf 17} (1977) 97\\
D.~Kazhdan, B.~Kostant and S.~Sternberg, Hamiltonian group action and dynamical
systems of Calogero type, Comm. on Pure and Appl.Math., Vol.XXXI (1978) 481-507

\bibitem{gn}
A.~Gorsky, N.~Nekrasov, Hamiltonian systems of Calogero type and two
dimensional YM theory, Nucl.Phys., {\bf B414} (1994) 213-23 \\
A.~Gorsky, N.~Nekrasov,
Relativistic Calogero-Moser systems as gauged WZW theory,
Nucl.Phys., {\bf B436} (1995) 582-608 \\
A.~Gorsky, N.~Nekrasov,
``Elliptic Calogero-Moser systems from two-dimensional current algebra'',
hepth/9401021


\bibitem{fgnr}
V.~Fock, A.~Gorsky, N.~Nekrasov and V.~Rubtsov,
``Duality in integrable systems and gauge theories'', JHEP, 0007, 028 (2000);
hep-th/9906235

\bibitem{bmmm}
H.~Braden, A.~Marshakov,A.~Mironov and A.~Morozov, hepth/9906240
``On Double-Elliptic Integrable Systems. 1. A Duality Argument for the case
of SU(2)", Nucl. Phys. {\bf B573}, 553-572, 2000. [hep-th/9906240]

\bibitem{HausTadd}
T.~Hausel,M.~Thaddeus ,
``Examples of Mirror Partners arising from Integrable Systems ",
C. R. Acad. Sci. Paris S\'er. I Math. , 333 (4) (2001) 313-318, math/0106140


\bibitem{gr}
A.~Gorsky and V.~Rubtsov,
``Dualities in integrable systems: Geometrical aspects,''
hep-th/0103004.
\bibitem{sklyanin}
E.~Sklyanin, Funk. Anal. Appl.,{\bf 16} (1982) 27

\bibitem{mukai}
S.~Mukai, ``Symplectic structure of the moduli space of sheaves on an abelian
or K3 surface'', Invent. Math. 77(1984), 101-116

\bibitem{OR}
A.~Odesskii, V.~Rubtsov,
``Polynomial Poisson Algebras with Regular Structure of
Symplectic Leaves'', ITEP-TH-42/01, math/0110032

\bibitem{Hitchin}
N.~Hitchin, Stable Bundles and Integrable Systems,
Duke Math. J.54(1987), no.1,91-114

\bibitem{Sz}
T.~Szemberg,
On principally polarized Adler-van Moerbeke surfaces,
Math.Nachr. {\bf 185}(1997), 239-260.

\bibitem{AvM}
M.~Adler, P.~van Moerbeke,
The intersection of four quadrics in ${\CC}P^6$, abelian surfaces and
their moduli, Math. Annalen, {\bf 278}(1987), 117-131.

\bibitem{Nekr}
N.~Nekrasov, Holomorphic bundles and many-body systems,
Comm.Math.Phys., {\bf 180} (1996) 587-604

\bibitem{ER}
B.~Enriquez, V.~Rubtsov, Math.Res.Lett. {\bf 3} (1996) 343

\bibitem{HurtMark}
J.~Hurtubise, E.~Markman,
math.AG/9912161, Calogero-Moser systems and Hitchin systems
\bibitem{aru}
G.~E.~Arutyunov, S.~A.~Frolov and P.~B.~Medvedev,
``Elliptic Ruijsenaars-Schneider model from the cotangent bundle over the
two-dimensional current group,'' J.\ Math .\ Phys.\  {\bf 38}, 5682 (1997)
hep-th/9608013.\\
G.~E.~Arutyunov, S.~A.~Frolov and P.~B.~Medvedev,
``Elliptic Ruijsenaars-Schneider model via the Poisson reduction of the
Affine Heisenberg Double,''
J.\ Phys.\ A {\bf A30}, 5051 (1997) hep-th/9607170\\

\bibitem{gnr}
A.~Gorsky, N.~Nekrasov and V.~Rubtsov, Hilbert schemes, separated variables and
D-branes, Comm. Math. Phys. {\bf 222} (2001) 299 ; hep-th/9901089

\bibitem{gm}
A.~Gorsky and A.~Mironov, ``Integrable many-body systems and gauge theories,''
hep-th/0011197.
\bibitem{BKr} ``Integrability: The Seiberg-Witten and Whitham Equations",
H.W. Braden and I.M. Krichever (editors), Gordon and Breach 2000.

\bibitem{nikita5}
N.~Nekrasov,
``Five dimensional gauge theories and relativistic integrable systems,''
Nucl.\ Phys.\ B {\bf 531}, 323 (1998)
[hep-th/9609219].
\bibitem{bmmm2} H. W. Braden, A. Marshakov, A. Mironov and A. Morozov,
``The Ruijsenaars-Schneider Model in the Context of Seiberg-Witten Theory,"
Nucl. Phys. {\bf B558}, 371-390, 1999. hep-th/9902205

\bibitem{ggm}
A.~Gorsky, S.~Gukov and A.~Mironov,
``SUSY field theories, integrable systems and their stringy/brane  origin. II,''
Nucl.\ Phys.\ B {\bf 518} (1998) 689
[hep-th/9710239].
\bibitem{csaki}
C.~Csaki, J.~Erlich, V.~V.~Khoze, E.~Poppitz, Y.~Shadmi and Y.~Shirman,
``Exact results in 5D from instantons and deconstruction,''
hep-th/0110188.

\bibitem{barth}
W.~Barth,
Abelian Surfaces with (1,2)-Polarization, Adv. Studies in Pure Math.,10
(1987), Algebraic Geometry, Sendai,1985, pp.41-84

\bibitem{LB}
H.~Lange, Ch.~Birkenhake,
 Complex Abelian Varieties, Springer Grund.
math.Wissenschaften, 302,1992, 435 p.

\bibitem{Beau}
A.~Beauville,
 Syst\`emes Hamiltoniens Compl\`etement Int\'egrables associ\'es aux
surfaces K3, Symposia Math. ,32 (1991), Acad.Press,  "Problems in the theory of
surfaces and their classifications", pp.25 -31.

\bibitem{Taka}
K.~Takasaki, ``Hyperelliptic Integrable Systems on K3 and Rational Surfaces,"
hepth/0007073

\bibitem{YZ}
S.T.~Yau, E.~Zaslow, BPS states, string duality and nodal curves on K3,
Nucl.Phys.  {\bf B}, 471(3)(1996), pp.503-512

\bibitem{minah}
J.~Minahan, D.~Nemeschansky, C.~Vafa, N.~Warner,
$E-$strings and $N=2$ Topological Yang-Mills Theories,
 hep-th/9802168
\bibitem{kriza}
I.~Krichever and A.~Zabrodin,
``Spin generalization of the Ruijsenaars-Schneider model, nonAbelian 2-d Toda chain and representations of Sklyanin algebra,''
Russ. Math. Surveys {\bf 50} (1995) 1101,
hep-th/9505039
\bibitem{gz}
A.~S.~Gorsky and A.~V.~Zabrodin, ``Degenerations of Sklyanin algebra
and Askey-Wilson polynomials,''
J.\ Phys.\ A {\bf 26}, L635 (1993) [hep-th/9303026]

\bibitem{hasegawa}
K.~Hasegawa, q-alg/9512029 \\
A.~Antonov, K.~Hasegawa and A.~Zabrodin,
``On trigonometric intertwining vectors and non-dynamical R-matrix for
 the Ruijsenaars model,'' Nucl.\ Phys.\ B {\bf 503}, 747
(1997), hep-th/9704074.

\bibitem{FO}
B.~Feigin, A.~Odesskii,
q-alg/9509021,'' Vector bundles on elliptic curve and Sklyanin algebras''
\bibitem{Donagi}
 R. Donagi, L. Ein and  R. Lazarsfeld,
``A Non-Linear Deformation of the Hitchin Dynamical System''
alg-geom/9504017
\bibitem{Fairlie}
D.~B.~Fairlie , ``An Elegant Integrable System,'' Phys.Lett. 119A, 438 (1987).

\bibitem{ward}
R.~Ward, J.Phys.A {\bf 20} (1987) 2679

\bibitem{audin}
M.~Audin, Spinning Tops, Cambridge University Press, 1996, 139 p.

\bibitem{nambu}
Y.~Nambu, ``Generalized Hamiltonian Dynamics,''
Phys. Rev.D {\bf 7}, 2405 (1973).

\bibitem{tacht}
G.~Dito, M.~Flato, D.~Sternheimer and L.~Takhtajan,
``Deformation quantization and Nambu mechanics,''
Commun.\ Math.\ Phys.\  {\bf 183}, 1 (1997) hep-th/9602016.

\bibitem{thooft}
G.~'t Hooft, ``Topology Of The Gauge Condition And New Confinement Phases In
Nonabelian Gauge Theories,'' Nucl.\ Phys.\ B {\bf 190}, 455 (1981).


\bibitem{bn}
H.~W.~Braden and N.~A.~Nekrasov, ``Space-time foam from non-commutative
instantons ,'' hep-th/991201 \\
H. W. Braden and N. A. Nekrasov, ``Instantons, Hilbert Schemes and
Integrability", in ``Dynamical Symmetries of Integrable
Quantum Field Theories", NATO ARW, Kiev 2000. [hep-th/0103204]

\bibitem{Ganor}
O.~J.~Ganor, A.~Y.~Mikhailov and N.~Saulina,
``Constructions of non commutative instantons on T**4 and K(3),''
hep-th/0007236.\\
Y.~E.~Cheung, O.~J.~Ganor, M.~Krogh and A.~Y.~Mikhailov, ``Instantons on a
non-commutative T(4) from twisted (2,0) and  little-string theories,''
Nucl.\ Phys .\ {\bf B564}, 259 (2000) hep-th/9812172.

\bibitem{freedman}
R.~Friedman, J.~W.~Morgan and E.~Witten,
``Vector bundles over elliptic fibrations,''
alg-geom/9709029.

\bibitem{Krich}
I.~Krichever,
``Vector bundles and Lax equations on algebraic curves'',
hep-th/0108110

\bibitem{LOZ}
A.~Levin, M.~Olshanetsky, A.~Zotov,
 "Hitchin Systems - symplectic maps and two-dimensional version",
nlin.SI/0110045.

\bibitem{ah}
N.~Arkani-Hamed, A.~G.~Cohen, D.~B.~Kaplan, A.~Karch and L.~Motl,
``Deconstructing (2,0) and little string theories,''
hep-th/0110146.
\end{thebibliography}
\end{document}